\documentclass[twocolumn,english,twocolumn,prl,showpacs]{revtex4}
\usepackage[latin9]{inputenc}
\usepackage{graphicx,color}
\usepackage{amssymb}
\usepackage{amsmath}
\usepackage[bookmarks]{hyperref}
\usepackage{babel}

\newcommand{\beq}{\begin{equation}}
\newcommand{\eeq}{\end{equation}}
\newcommand{\be}{\begin{equation}}
\newcommand{\ee}{\end{equation}}

\begin{document}

\title{The Crooks relation in optical spectra - \\ universality in work distributions for weak local quenches}

\author{M. Heyl}
\affiliation{Department of Physics, Arnold Sommerfeld Center for Theoretical Physics,
and Center for NanoScience, Ludwig-Maximilians-Universit\"at M\"unchen,
Theresienstr. 37, 80333 Munich, Germany}
\author{S. Kehrein}
\affiliation{Department of Physics, Arnold Sommerfeld Center for Theoretical Physics,
and Center for NanoScience, Ludwig-Maximilians-Universit\"at M\"unchen,
Theresienstr. 37, 80333 Munich, Germany}
\affiliation{Georg-August-Universit\"at G\"ottingen, Friedrich-Hund-Platz 1, 37077 G\"ottingen}

\begin{abstract}
	We show that work distributions and non-equilibrium work fluctuation theorems can be measured in optical spectra for a wide class of quantum systems. We consider systems where the absorption or emission of a photon corresponds to the sudden switch on or off of a local perturbation. For the particular case of a weak local perturbation, the Crooks relation establishes a universal relation in absorption as well as in emission spectra. Due to a direct relation between the spectra and work distribution functions this is equivalent to universal relations in work distributions for weak local quenches. As two concrete examples we treat the X-ray edge problem and the Kondo exciton.
\end{abstract}

\pacs{05.40.-a,78.70.Dm,78.67.Hc}

\maketitle

Equilibrium thermodynamics provides the framework for the description of the equilibrium properties of macroscopically large systems. This includes the properties of systems in equilibrium states as well as the description of transitions between different equilibrium states even if the system is not in equilibrium in the meantime. Starting in 1997 with a seminal contribution from Jarzynski~\cite{Jarzynski}, the field of non-equilibrium work fluctuation theorems~\cite{Campisi_Rev} opened up. These relate a measurable non-equilibrium quantity, the work performed, to equilibrium free energies even if the system is driven arbitrarily far away from equilibrium.

Suppose a system is prepared in a thermal state at inverse temperature $\beta$. If the Hamiltonian $H(t)$ of the system changes during a time interval from $0$ to $t_f$ according to a prescribed protocol, work is performed on the system. In order to determine the work done two energy measurements are necessary leading to the notion that work is not an observable~\cite{Talkner_work}; the work $W$ rather is a random variable with a probability distribution function~\cite{Talkner_work}
\beq
	P_F(W)=\int \frac{ds}{2\pi} \: e^{iWs} G(s), \:\: G(s)=\left\langle e^{i H(0) s} e^{-i H_H(t_f) s} \right\rangle.
\label{eq_def_P}
\eeq
Here $\langle \cdots \rangle$ denotes the thermal average over the initial state and $H_H(t_f)=U^\dag(t_f) H(t_f) U(t_f)$ with $U(t_f)$  the time-evolution operator obeying the differential equation $i\partial_t U(t)=H(t) U(t)$. In this Letter we set $\hbar=1$.

Let $P_{B}(W)$ be the probability distribution function for the backward protocol. Then the Crooks relation, first shown for classical systems~\cite{Crooks} and later extended to closed as well as open quantum systems~\cite{Tasaki,Talkner_Crooks,Talkner_Crooks_open}:
\beq
	\frac{P_{F}(W)}{P_{B}(-W)}=e^{\beta(W-\Delta F)},
\label{Crooks_relation}
\eeq
establishes a universal connection between the forward and backward processes that only depends on the equilibrium free energy difference $\Delta F$ of the final and initial state independent of the details of the protocol. The Jarzynski relation~\cite{Jarzynski} is a consequence of Eq.~(\ref{Crooks_relation}), see e.g. Ref.~\cite{Crooks}.

Experimental tests of the Crooks relation have been performed in recent years for classical systems. Among these are folding-unfolding experiments of small RNA-hairpins where the free energy difference between the folded and unfolded state has been 	extracted using the Crooks relation~\cite{Collin,Junier}. Moreover, it has been verified in electrical circuits~\cite{Garnier}, for mechanical oscillators~\cite{Douarche}, small colloidal particles ~\cite{Wang} and nonthermal systems~\cite{Schuler}.

In the quantum case a measurement of work distributions has not been performed up to now. Recently, a measurement scheme in optical traps has been proposed~\cite{Huber} that has not been realized yet. In the present Letter we show that work distributions of quantum systems have been measured for decades in terms of X-ray spectra of simple metals. We point out that there exists a large class of quantum systems associated with the X-ray edge problem where absorption spectra $A(\omega)$ and emission spectra $E(\omega)$ can be identified with forward and backward work distributions for a sudden switch on or off of a local perturbation. This allows for an experimental observation of non-equilibrium work fluctuation theorems such as the Crooks relation. For the particular case of a weak local perturbation, the Crooks relation manifests in the universal relations
\beq
	\frac{A(\omega+\Delta F)}{A(-\omega+\Delta F)}=e^{\beta \omega},\:\:\frac{E(\omega+\Delta F)}{E(-\omega+\Delta F)}=e^{-\beta \omega}
\label{eq_Crooks_Abs}
\eeq
that hold in second order renormalized perturbation theory. This is the central result of this Letter that will be proven below. Here $\Delta F$ is the free energy difference between the system with and without local perturbation at the same inverse temperature $\beta$. Notice that an independent measurement of $\Delta F$ is not required to establish Eq.~(\ref{eq_Crooks_Abs}) in an experiment. Actually, Eq.~(\ref{eq_Crooks_Abs}) permits a determination of $\Delta F$ similar to experiments in biophysics~\cite{Collin,Junier}. Due to the correspondence between spectra and work distributions, Eq.~(\ref{eq_Crooks_Abs}) implies universal relations for work distributions of weak local quenches:
\beq
	 \frac{P_F(W+\Delta F)}{P_F(-W+\Delta F)}=e^{\beta W},\:\:\frac{P_B(W-\Delta F)}{P_B(-W-\Delta F)}=e^{\beta W}.
\label{eq_univ_work}
\eeq
Here, $P_F(W)$ is the work distribution for a protocol where the local perturbation is suddenly switched on and $P_B(W)$ the work distribution for the backward protocol.

Consider a system weakly coupled to a monochromatic light field of frequency $\omega$ where the absorption or emission of a photon corresponds to the sudden switch on or off of a local perturbation. Such systems have been discussed extensively in the literature. In the X-ray spectra of simple metals a system of free fermions has to adapt to a suddenly created or annihilated local potential scatterer~\cite{Mahan,Noziere,Schotte,X_Rev}. For metals with incomplete shells the local perturbation is related to localized orbitals~\cite{K_FES,X_Rev}. As has been shown recently, spectra of quantum dots allow for an idealized implementation of X-ray edge type problems~\cite{Heyl_QD,Tureci}. In the remainder, $H$ denotes the Hamiltonian with the local perturbation and $H_0$ without, respectively.

\emph{Crooks relation in absorption and emission spectra.} First, we concentrate on the absorption case, the related emission spectra will be discussed below. The absorption spectrum for incident light of frequency $\omega$ in second order of the system-light field coupling (Fermi's golden rule) is related to a dynamical correlation function via Fourier transformation
\beq
	A(\omega)=\kappa_A \int \frac{dt}{2\pi} \: e^{i \omega t} \: G_A(t).
\label{eq_A_FT}
\eeq
The constant $\kappa_A$ contains parameters depending on the experimental details such as the intensity of the incident light beam or the system-light field coupling. Note that the photon energy $\omega$ in Eq.~(\ref{eq_A_FT}) is not the bare one, it is usually measured relative to a constant offset $\omega_o$, e.g. the core-hole binding energy in the x-ray edge problem. We consider those systems where the dynamical correlation function $G_A(t)$ appearing in Eq.~(\ref{eq_A_FT}) is of the structure
\beq
	G_A(t)=\frac{1}{Z_A} \textrm{Tr}  \left( e^{-\beta H_0} e^{i H_0 t} e^{-i H t} \right), \:\: Z_A=\textrm{Tr} \left( e^{-\beta H_0} \right)
\label{eq_G_A}
\eeq
as in the case of X-ray edge type problems ~\cite{Mahan,Noziere,Schotte,K_FES,X_Rev,Heyl_QD,Tureci}. For a particular problem at hand, the question wether $G_A(t)$ can be brought into the form in Eq.~(\ref{eq_G_A}) has to be studied on a case by case basis. Regarding Eq.~(\ref{eq_G_A}) $G_A(t)$ is the characteristic function of a work distribution for a quench from $H_0$ to $H$, cf. Eq.~(\ref{eq_def_P}). This identification allows for an observation of the Crooks relation in an optics experiment. Recently, X-ray edge singularities have been found in work distributions for local quenches in an Ising chain at criticality~\cite{Silva}.

The emission spectrum $E(\omega)$ corresponding to the same setup is given by
\beq
	E(\omega)=\kappa_E \int \frac{dt}{2\pi} \: e^{-i \omega t} G_E(t)
\eeq
with
\beq
	G_E(t)=\frac{1}{Z_E} \textrm{Tr} \left( e^{-\beta H} e^{i H t} e^{-i H_0 t} \right), \: \: Z_E=\textrm{Tr} \left( e^{-\beta H} \right).
\label{eq_G_E}
\eeq
Hence, $E(-\omega)$ is proportional to the work distribution for a protocol where the local perturbation is switched off, that is precisely the backward process to absorption. A direct application of the Crooks relation in Eq.~(\ref{Crooks_relation}) therefore yields
\beq
	\frac{A(\omega)}{E(\omega)}=\frac{\kappa_A}{\kappa_E} e^{\beta(\omega-\Delta F)}
\label{eq_Crooks_Abs_Em}
\eeq
as an exact result. This relation depends on experimental details through the parameters $\kappa_A$ and $\kappa_E$. The linear scaling of $\ln(A(\omega)/E(\omega))$ as a function of the frequency $\omega$ of the light beam, however, is universal with a slope $\beta$. Note that Eq.~(\ref{eq_Crooks_Abs_Em}) is valid for an arbitrary strength of the local perturbation, we only assume a small coupling to the external light field.

Two different measurements, absorption and emission, are necessary to explore this relation in experiment. However, the Crooks relation can also be measured in a single experiment in case of weak local perturbations where Eq.~(\ref{eq_Crooks_Abs}) holds as will be shown below. This has the additional advantage as opposed to the exact relation in Eq.~(\ref{eq_Crooks_Abs_Em}) that also the experiment specific constants $\kappa_A$ and $\kappa_E$ drop out.

Eqs.~(\ref{eq_def_P},\ref{eq_G_A},\ref{eq_G_E}) show the formal equivalence between work distribution functions and optical x-ray edge spectra. 
In conventional experiments the work distribution function is sampled by recording in each realization the work performed. The full distribution function is successively built up jointly over all work values.
Optical spectra, however, are recorded differently. The outcome of a measurement is not the work performed. Instead one obtains directly the probability for photon absorption (or emission) at a given frequency $\omega$ (work performed). The full distribution function is then constructed by sweeping the laser through all relevant frequencies. 

The advantage of measuring work distributions via optical spectra is that the absorbed photon carries out the sequence of measuring the energy in the initial state, applying the perturbation, and measuring the energy of the final state, in a single step. It can be absorbed or emitted only in case when its frequency $\omega$ matches precisely the energy difference between the system's initial and final state. The disadvantage is that only specific local perturbations and only specific protocols (sudden switchings) can be implemented.

\emph{Crooks relation in a single spectrum.} Suppose $V$ is the unitary transformation that diagonalizes the Hamiltonian $H$. In the following we normal order Hamiltonians relative to the finite temperature initial mixed state~\cite{Wick}. For generic weak coupling impurity problems the diagonalized Hamiltonian can be represented as~\cite{Supplementary_information}
\begin{equation} 
	VHV^\dag=H_0+\Delta F
\label{diag_ident}
\end{equation}
in the thermodynamic limit~\cite{bound_states} where $\Delta F$ denotes the free energy difference between the systems described by $H$ and $H_0$ at the same temperature $T$. The appearance of temperature in this equation can be understood from the normal ordering procedure~\cite{Supplementary_information}. As a consequence of Eq.~(\ref{diag_ident}), the dynamical correlation functions $G_{A/E}(t)$ can be written as:
\begin{eqnarray}
	G_A(t) &=&\frac{1}{Z_A} \textrm{Tr} \left( e^{-\beta H_0} V^\dag(t) V \right) e^{-i\Delta F t}, \nonumber  \\ G_E(t)&=&\frac{1}{Z_A} \textrm{Tr} \left( e^{-\beta H_0} V(t) V^\dag \right)e^{i\Delta F t} e^{-\beta \Delta F}
\label{eq_G_U}
\end{eqnarray}
where $V(t)=e^{iH_0 t} V e^{-i H_0 t}$ and $\Delta F=-\beta^{-1}\log(Z_E/Z_A)$. For all the relevant cases, it is possible to represent the unitary transformation $V$ as an ordered exponential $V=\textrm{O} \exp[\chi]$ where $\chi$ is antihermitian, $\chi^\dag=-\chi$, and $\textrm{O}$ denotes some ordering prescription. For generic weak coupling problems such as the Kondo model at nonzero temperature analyzed later, the flow equation approach provides a general prescription for the construction of the unitary transformation $V$ as an ordered exponential of its generator $\eta(B)$~\cite{Flow_equation}
\beq
	V=\mathcal{T}_B \exp \left[ \int_0^\infty dB \: \eta(B) \right]
\label{eq_U_feq}
\eeq
where $\eta(B)$ is determined by a set of differential equations. For $B > B'$, $\mathcal{T}_B$ orders an $\eta(B)$ left of an $\eta(B')$. Expectation values of ordered exponentials such as in Eq.~(\ref{eq_G_U}) can be related to the exponential of a cumulant average~\cite{Kubo} that can be expanded in a power series in powers of $\chi$. The first cumulant vanishes as $\chi$ can be chosen normal ordered relative to the initial state. For the X-ray edge problem the cumulant expansion stops at second order within the validity of the bosonization technique, see below. For more complicated problems such as the Kondo exciton the diagonalizing unitary transformation can be obtained by the flow equation framework, see Eq.~(\ref{eq_U_feq}). In this case, the generator $\eta(B)$ and thus the operator $\chi$ is proportional to the strength of the local perturbation such that in the case of a weak local perturbation the expansion is controlled by a small parameter. For systems with significant renormalization effects, couplings have to stay small over the whole renormalization flow.

Performing this cumulant expansion up to second order one observes that $G_A(t)$ and $G_E(t)$ are directly related to each other via $G_A(t)e^{i\Delta F t}=G_E(t)e^{-i \Delta F t}e^{-\beta \Delta F}$. For the spectra this result implies $\kappa_A E(\omega+\Delta F) e^{-\beta \Delta F}=\kappa_E A(-\omega+\Delta F)$.
Plugging this relation into the Crooks relation, see Eq.~(\ref{eq_Crooks_Abs_Em}), one directly proves the main result, Eq.~(\ref{eq_Crooks_Abs}), in second order renormalized perturbation theory.

In the remainder of this Letter, we will discuss two examples for the Crooks relation in absorption spectra: the X-ray edge problem and the Kondo exciton.

\emph{The X-ray edge problem.} In the X-ray edge problem the absorption of a photon is accompanied by the sudden creation of a local potential scatterer in a sea of noninteracting fermions~\cite{Noziere}. Hence, we have $H_0=\sum_{k} \varepsilon_k \colon c_k^\dag c_k \colon $ and $H=H(g)=H_0+(2 \pi /L) g\sum_{kk'} \colon c_k^\dag c_{k'}\colon$ where the colons denote normal-ordering, see \cite{Supplementary_information}. We consider a linearized dispersion $\varepsilon_k=v_F k$ and set $v_F=1$. The Fourier transform of the absorption spectrum is given by~\cite{Noziere}
\beq
	S(t)=\frac{1}{Z_A} \textrm{Tr} \left( e^{-\beta H_0} e^{i H_0 t} \psi(0) e^{-i H(g) t} \psi^\dag(0) \right)
\label{eq_G_A_X_ray}
\eeq
that is yet not in the desired form as in Eq.~(\ref{eq_G_A}). Using the bosonization technique, the fermionic fields $\psi(x)$ can be represented in terms of bosonic ones, $\phi(x)$, via $\psi(x)=a^{-1/2}F e^{-i\phi(x)}$ with $a^{-1}$ an ultra-violet cutoff~\cite{Schoeller_Delft}. The Klein factor $F$ commutes with $H(g)$ and does not contribute to $S(t)$ due to its property $FF^\dag=1$. The bosonization identity allows to regard the fermionic fields as a unitary transformation acting on $H(g)$ such that $S(t) \propto G_A(t)e^{-i \Delta t}$ with a constant energy shift $\Delta$ that can be absorbed into a redefinition of the constant offset $\omega_o$ and $G_A(t)$ is in the desired form:
\beq
	G_A(t)=\frac{1}{Z_A} \textrm{Tr} \left( e^{-\beta H_0} e^{i H_0 t} e^{-i H(1+g) t} \right).
\eeq
The diagonalizing transformation $V$ of $H(1+g)$ equals $V=e^{i (1+g) \phi(0)}$~\cite{Schotte}. Although the effective strength of the scatterer $1+g$ is not small, the cumulant expansion stops at second order as the operator in the exponent is linear in bosonic operators. Hence, in the range of validity of the bosonization treatment, the Crooks relation in Eq.~(\ref{eq_Crooks_Abs}) holds exactly for the X-ray edge absorption spectrum. Comparing bosonization~\cite{Schotte} with the exact treatment~\cite{Noziere}, it yields the correct result up to second order in $g$. This restriction originates from the linearization of the free fermionic spectrum~\cite{Schotte}.

\begin{figure}
\centering
\includegraphics[width=\columnwidth]{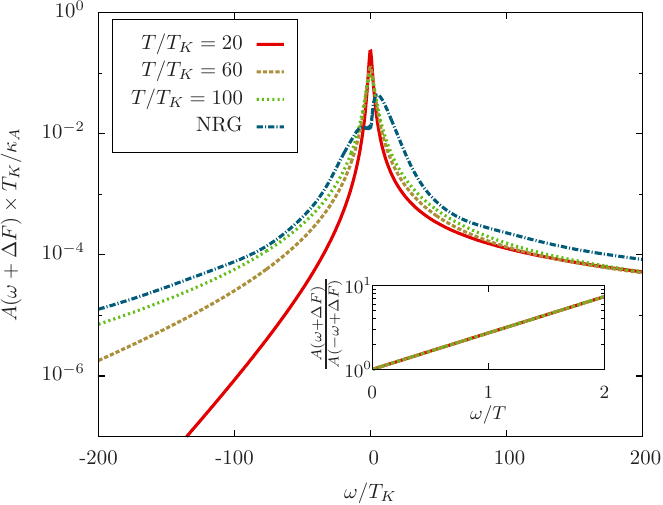}
\caption{Absorption spectrum of a Kondo exciton for different temperatures as a function of the light frequency. As a reference, a NRG curve for $T=100 \, T_K$ is shown taken from T\"ureci \emph{et al.}~\cite{Tureci}. In the regime $|\omega|\gtrsim T$ where the NRG data is accurate the agreement is excellent within the numerical accuracy. The inset exemplifies the validity of the Crooks relation in the absorption spectrum of the Kondo exciton, all curves for $T/T_K=20,60,100$ lie on top of each other when plotted against $\omega/T$.}
\label{Fig1}
\end{figure}

\emph{The Kondo exciton.} Recently, T\"ureci \emph{et al.}~\cite{Tureci} proposed an experimental setup for a quantum dot where the absorption of a photon corresponds to the sudden switch on of a Kondo impurity. Hence, we have $H_0=\sum_{k\sigma} \varepsilon_k \colon c_{k\sigma}^\dag c_{k \sigma }\colon$ and $H=H_0+\sum_{kk'}J_{kk'} \colon\vec{S} \cdot  \vec{s}_{kk'}\colon$. For details about the Kondo problem see for example Ref.~\cite{Hewson}. The dynamical correlation function $G_A(t)$ for the absorption spectrum is given by Eq.~(\ref{eq_G_A}). The diagonalizing unitary transformation $V$ can be obtained by the flow equation approach~\cite{Flow_equation}, cf. Eq.~(\ref{eq_U_feq}), with $\eta(B)=\sum_{kk'} \left( \varepsilon_k-\varepsilon_{k'} \right) J_{kk'}(B) \colon \vec{S} \cdot \vec{s}_{kk'}\colon$ in 1-loop order. The couplings $J_{kk'}(B)$ are determined by a set of differential equations~\cite{Peter_Stefan}. Importantly, the flow equation framework includes all the renormalization effects such as the emergence of a low-energy scale $T_K$, the Kondo temperature. The absorption spectrum is obtained via the cumulant expansion up to second order in the coupling strength. Its validity is restricted to weak coupling problems such that we have to require $T\gg T_K$~\cite{details}. A plot of the absorption spectrum is shown in Fig.~\ref{Fig1} for different temperatures. As a reference, a NRG-curve for $T=100 \, T_K$ obtained by T\"ureci \emph{et al.}~\cite{Tureci} for an Anderson impurity model in the Kondo regime is included in this figure~\cite{TK}. In the vicinity of the main peak at small $|\omega|<T$, the NRG calculation contains an unphysical double peak structure. For more details we refer to Ref.~\cite{Tureci}. For frequencies $|\omega|\gtrsim T$, however, where the NRG data is accurate we observe excellent agreement with the results of the flow equation formalism. Asymptotic formulas for $A(\omega)$ in the limit $\omega\to\pm\infty$ can be found in Ref.~\cite{Tureci}. The inset shows the validity of  Eq.~(\ref{eq_Crooks_Abs}). The ratio $A(\omega+\Delta F)/A(-\omega+\Delta F)$ is the universal function $e^{\beta \omega}$ independent of any details.

\emph{Conclusions.} We have shown that work distributions and thus non-equilibrium work fluctuation theorems can be measured in optical spectra of quantum systems such as the X-ray edge problem or the Kondo exciton. For weak local perturbations, the Crooks relation establishes a universal relation within a single spectrum, absorption or emission, cf. Eq.~(\ref{eq_Crooks_Abs}). 

\emph{Acknowledgements.} We acknowledge fruitful discussions with Jan~von~Delft and Peter H\"anggi. We thank Markus Hanl and Andreas Weichselbaum for providing us the
NRG curve in Fig.~\ref{Fig1}. This work was supported by SFB~TR12 of the Deutsche Forschungsgemeinschaft (DFG), the Center for Nanoscience (CeNS) Munich, and the German Excellence Initiative via the Nanosystems Initiative Munich (NIM).

\bibliographystyle{apsrev}

\appendix*

\section{Supplementary information: Details about the diagonal structure of the locally perturbed Hamiltonian}

In this supplementary information we want to analyze the general statement about the diagonal structure of locally perturbed Hamiltonians in Eq.~(10) of the main text for the case of a paradigmatic example, the potential scattering Hamiltonian $H_p$:
\be
	H_p = H_0 + g \frac{2\pi}{L} \sum_{kk'} c_k^\dag c_{k'}, \quad H_0 = \sum_{k} \varepsilon_k c_k^\dag c_k .
\label{eq:H_p}
\ee
The fermionic operator $c_k^\dag$ creates a fermion in a state with wave vector $k$. For simplicity we restrict to the case of spinless fermions. The parameter $g$ is the scattering amplitude and $L$ is the system size. 

\subsection{Normal-ordering and energy offset}

In the main text we normal-ordered the Hamiltonians relative the initial finite temperature mixed state at inverse temperature $\beta$. In the following, the normal ordered counterparts will be denoted by an additional superscript $^n$
\begin{align}
	H_0^n  = &  H_0 - \langle H_0 \rangle_0, \nonumber \\ & \langle \dots \rangle_0 = Z_0^{-1} \mathrm{Tr} \left[ e^{-\beta H_0} \dots \right], \nonumber \\ & Z_0 = \mathrm{Tr}  \left[ e^{-\beta H_0} \right] .
\label{eq:H_pn}
\end{align}
In comparison with Eq.~(\ref{eq:H_p}) the normal-ordered Hamiltonian $H_0^n$ is shifted by a constant energy contribution $\langle H_0 \rangle_0$. In case of nonquadratic interacting systems such as the Kondo model in the main text the appropriate normal-ordering prescription is given by Wick~\cite{Wick}. 

In principle, temperature is a statistical property emerging on the macroscopic level, thus, it should not appear in the microscopic description of the system in terms of the Hamiltonian. However, there is a principal freedom of fixing the arbitrary global energy offset. In this work we choose the global offset in such a way that the thermal expectation value of the unperturbed system $\langle H_0^n \rangle_0=0$ vanishes in the initial state. The corresponding shifted potential scattering Hamiltonian is then given by
\be
	\tilde{H}_p = H_p - \langle H_0 \rangle_0
\ee

\subsection{Diagonalizing transformation}

The potential scattering Hamiltonian in Eq.~(\ref{eq:H_p}) can be diagonalized via a unitary transformation $V$
\begin{align}
	H_p^d & = V H_p V^\dag = \sum_{k} E_k c_k^\dag c_k, \nonumber \\ & V=\exp\left[ \chi \right], \quad \chi = \sum_{kk'} v_{kk'} c_k^\dag c_{k'},
\label{eq:diag_trafo}
\end{align}
for a suitably chosen matrix $v_{kk'}$ that is proportional to $g$ for weak potential scatterers. As the potential scatterer is a local perturbation in a large system of size $L$ the dispersion $E_k$ coincides with $\varepsilon_k$ up to finite-size corrections
\be
	E_k = \varepsilon_k + \frac{\pi}{L} \delta_k.
\label{eq:dispersion}
\ee
Thus $\lim_{L\to \infty} E_k=\varepsilon_k$. Note that we restrict to cases without bound states where contributions to the spectrum of $\mathcal{O}(1)$ can appear instead of $\mathcal{O}(L^{-1})$ as above. Concerning global system properties, the finite-size corrections given by $\delta_k$ vanish in the thermodynamic limit, the local system properties, however, are determined by the $\delta_k$'s. Regarding the time evolution of the single-particle operators 
\be
	e^{i H_p^d t} c_k e^{-i H_p^d t} \stackrel{L \to \infty}{\longrightarrow}c_k e^{-i \varepsilon_k t}
	\label{eq:time_evol_ck}
\ee
the finite-size effects can be neglected in the thermodynamic limit as we always have $L \gg t$. The unitary transformation $V$ not only diagonalizes $H_p$ but at the same time also $\tilde{H}_p$ such that
\begin{align}
	V \tilde{H}_p V^\dag & = H_p^d  -  \langle H_0 \rangle_0 = \nonumber \\ & = H_0^n + \frac{\pi}{L} \sum_k \delta_k \colon c_k^\dag c_k \colon + \frac{\pi}{L} \sum_k \delta_k f(\varepsilon_k)
\label{eq:diag_Hptilde}
\end{align}
Here, the colons $\colon \dots \colon$ denote normal-ordering which for the present quadratic Hamiltonian is a short-hand version for $\colon c_k^\dag c_k \colon = c_k^\dag c_k - \langle c_k^\dag c_k \rangle_0$.

The last term in the above equality is the free energy difference $\Delta F$ between the system with and without local perturbation
\begin{align}
	\Delta F & = - \beta^{-1} \sum_k \log\left[ 1 + e^{-\beta E_k} \right] + \nonumber \\ & + \beta^{-1} \sum_k \log\left[ 1 + e^{-\beta \varepsilon_k} \right] = \nonumber \\ & =  \frac{\pi}{L} \sum_k \delta_k f(\varepsilon_k) + \mathcal{O}(L^{-1})
\label{eq:DeltaF}
\end{align}
with $f(\varepsilon_k) = 1/[1+e^{\beta \varepsilon_k}]$ the Fermi-Dirac distribution. Note that for the internal energy difference $\Delta U$ we have $\Delta U = \sum_k E_k f(E_k) - \sum_k \varepsilon_k f(\varepsilon_k) = \Delta F +  (\pi/L) \sum_k \delta_k \varepsilon_k f'(\varepsilon_k)$ with $f'$ the derivative of the Fermi-Dirac distribution.

The second term on the right hand side of Eq.~(\ref{eq:diag_Hptilde}) denotes a finite-size correction to $H_0^n$. This correction vanishes when evaluated in the initial finite temperature Gibbs state. Concerning the time evolution of the single-particle operators the corresponding contribution vanishes in the thermodynamic limit, see Eq.~(\ref{eq:time_evol_ck}). 
As we will show below in case of the absorption spectrum for the potential scattering Hamiltonian this part of the Hamiltonian gives no contribution at all. Note that this property strongly depends on the initial state. For the remainder of this supplementary information we will thus use the following identity
\be
	V H_p^n V^\dag \stackrel{L\to \infty}{\longrightarrow} H_0^n +\Delta F.
\label{eq:Hpd_DeltaF}
\ee

\section{Absorption spectrum}

In the following, we show that the approximate identity in Eq.~(\ref{eq:Hpd_DeltaF}) yields precisely the absorption spectrum one obtains from the exact solution.

Based on the knowledge of the diagonalizing transformation $V$ one can calculate the Fourier transform $G_A(t)$ of the absorption spectrum, see Eq.~(6) in the main text, also without the identity in Eq.~(\ref{eq:Hpd_DeltaF}). For $G_A(t)$ we have
\begin{align}
	G_A(t) & = Z_A^{-1} \mathrm{Tr} \left[ e^{-\beta H_0} e^{-i (H_p^d- H_0) t} V^\dag(t) V \right], \nonumber \\ & V^\dag(t) = e^{i H_0 t} V^\dag e^{-i H_0 t}.
\end{align}
where we have used that the finite-size effects for the time evolution can be neglected, see Eq.~(\ref{eq:time_evol_ck}). The expressions for the emission spectrum can be obtained analogously. Using the exponential representation of $V$ in Eq.~(\ref{eq:diag_trafo}) we perform a cumulant expansion in the expansion parameter $g$ leading to
\begin{align}
	G_A(t) & = \exp\left[ -i  \langle H_p^d - H_0 \rangle_0^c \: t + \langle \chi - \chi(t) \rangle_0^c + \right. \nonumber \\ & \left. \frac{1}{2}\langle \chi \chi - \chi(t)\chi  \rangle_0^c + \mathcal{O}(g^3)\right]
\end{align}
because $\chi$ is proportional to $g$ for weak potential scatterers. The superscript $^c$ indicates a cumulant average. All higher cumulants $\langle (H_p^d-H_0)^n \rangle_0^c$ for $n>1$ vanish in the thermodynamic limit. The same is true for the mixed cumulants $\langle (H_p^d-H_0)^n \chi^m \rangle_0^c$ for $n,m>0$. As $\langle \chi(t) \rangle_0=\langle \chi \rangle_0$ the lowest order contribution is of second order in $g$. The first term is the free energy difference as in Eq.~(\ref{eq:DeltaF})
\be
	\langle H_p^d -H_0\rangle_0 = \frac{\pi}{L}\sum_k \delta_k f(\varepsilon_k) = \Delta F.
\ee
Concluding one obtains
\be
	G_A(t)=\exp\left[ -i \Delta F t +\frac{1}{2} \langle \chi \chi -\chi(t) \chi \rangle_0 \right].
\label{eq:GA}
\ee
Using normal-ordered Hamiltonians and Eq.~(\ref{eq:Hpd_DeltaF}) the absorption spectrum is determined by the generating function
\be
	G_A(t) = Z_A^{-1} \mathrm{Tr} \left[e^{-\beta H_0^n} e^{i H_0^n t} e^{-i \tilde{H}_p t} \right].
\ee
Based on Eq.~(\ref{eq:Hpd_DeltaF}) one can write
\be
	G_A(t) = Z_A^{-1} \mathrm{Tr} \left[ e^{-\beta H_0^n} V^\dag(t) V\right] e^{-i \Delta F t}
\ee
with $V^\dag(t)=e^{i H_0 t} V^\dag e^{-i H_0 t}$ as before. Performing again the cumulant expansion one obtains the same result as in Eq.~(\ref{eq:GA}).

\bibliographystyle{apsrev}

\begin{thebibliography}{30}
\bibitem{bound_states} We restrict to situations without bound states.
\bibitem{details} Details of this calculation will be presented elsewhere.
\bibitem{TK} There is no fitting parameter since the Kondo temperatures from the $T=0$ impurity contribution to the specific heat ($T_K^{NRG}$) and of the high temperature expansion via the flow equations ($T_K^{feq}$) are linked by the Wilson number $w=0.4128$ via $T_K^{feq}=w T_K^{NRG}$~\cite{Hewson}. For the curves in Fig.~\ref{Fig1} the temperatures are chosen according to $T_K^{NRG}$, i.e., $T_K=T_K^{NRG}$.
\end{thebibliography}

\end{document}